\documentclass[11pt, preprintnumbers, amsmath,amssymbre]{revtex4}
\usepackage{amsmath}
\usepackage{amsfonts}
\usepackage{amssymb}
\usepackage[colorlinks,linkcolor=blue,citecolor=red]{hyperref}
\usepackage{graphicx}
\usepackage{hyperref}
\usepackage{subfigure}
\usepackage{graphicx}
\usepackage{dcolumn}
\usepackage{bm}
\usepackage{amsmath,amsthm,amsfonts}
\usepackage{graphicx}
\usepackage{physics}
\usepackage{hyperref}
\usepackage{lscape} 
\usepackage[pdf]{pstricks}
\usepackage{mathtools}
\usepackage{appendix}
\usepackage{comment}
\usepackage{pdflscape}
\graphicspath{{figs/}}

\newcommand{\nc}{\newcommand}
\nc{\ba}{\begin{eqnarray}}
\nc{\ea}{\end{eqnarray}}
\nc{\bfk}{\bf{k} }
\nc{\bfq}{\bf{q} }
\nc{\rc}{\textcolor[rgb]{1.00,0.00,0.00}}
\nc{\bc}{\textcolor[rgb]{0.00,0.07,1.00}}

\begin{document}

\title{  Vacuum zero point energy of self-interacting quantum fields in dS background}


\author{Hassan Firouzjahi$^{a}$}
\email{firouz@ipm.ir}

\author{Haidar Sheikhahmadi$^{a}$}
 \email{h.sh.ahmadi@gmail.com;h.sheikhahmadi@ipm.ir}

\affiliation{$^{a}$ School of Astronomy, Institute for Research in Fundamental Sciences (IPM),  P. O. Box 19395-5531, Tehran, Iran}

\begin{abstract}

We consider self-interacting scalar fields with a conformal coupling in the dS background and study the quantum corrections from bubble loop diagrams.  Incorporating the perturbative in-in formalism, we calculate the quantum corrections in the vacuum zero point energy and pressure of self-interacting fields with the potential  $V \propto \Phi^n $ for even values of $n$. We calculate the equation of state corresponding to these quantum corrections and examine the scaling of the divergent terms in the vacuum zero point energy and pressure associated to  the dimensional regularization scheme.
In particular, we show that for  quartic self-interacting scalar field the conformal invariance is respected at two-loop order at the conformal point.

\end{abstract}

\maketitle


\section{introduction}
\label{Intro00}

Quantum field theory in dS background is a rich topic which has important implications both observationally  and theoretically  \cite{DeWitt:1975ys, Birrell:1982ix, Fulling:1989nb, Mukhanov:2007zz, Parker:2009uva}.
Observationally, there are strong evidences that the early universe experienced a period of inflation in which the background was nearly a dS spacetime. In the simplest realization, inflation is driven by a scalar field, the inflaton field, with a near flat potential \cite{Weinberg:2008zzc, Baumann:2022mni}. While the inflaton field slowly rolls on top of its potential, its quantum fluctuations are stretched on superhorizon scales, which provide the seeds of large scale structure and the perturbations on CMB \cite{Kodama:1984ziu, Mukhanov:1990me}. The basic predictions of the models of inflation are that these primordial perturbations are nearly scale invariant, Gaussian and adiabatic which are well supported by cosmological observations \cite{Planck:2018vyg, Planck:2018jri}. In addition,
numerous  observations indicate that the late universe is undergoing a phase of accelerating expansion.  The origin of dark energy as the source of the recent cosmological acceleration  is not known but a cosmological constant associated with the quantum zero point energy of fields may be a good option  \cite{Weinberg:1988cp, Sahni:1999gb, Peebles:2002gy, Copeland:2006wr, Martin:2012bt}.

On the theoretical sides, while there is no compelling theory of quantum gravity at hand, but  understanding quantum fields in curved backgrounds, including the dS background, may shed some lights for the pursuit of   a  theory of quantum gravity. Understanding important issues  such as regularizations and renormalization of cosmological correlations of quantum perturbations  in dS background can play important roles  as well.
More specifically,  similar to  quantum field theories in flat spacetime, physical quantities such as the energy momentum tensor, energy density and pressure suffer from infinities in a curved spacetime.  Therefore, it is an important question as how one can regularize and renormalize  the infinities to  extract the finite physical quantities.
Furthermore, the fact that there is no unique vacuum in a curved spacetime adds more complexities for the treatment of regularization and renormalization in a curved spacetime \cite{Hawking:1975vcx, Unruh:1976db, Unruh:1983ms, Jacobson:2003vx, Cozzella:2020gci, Firouzjahi:2022rtn}.

\begin{figure}[t]
	\centering
	\includegraphics[ width=1.0\linewidth]{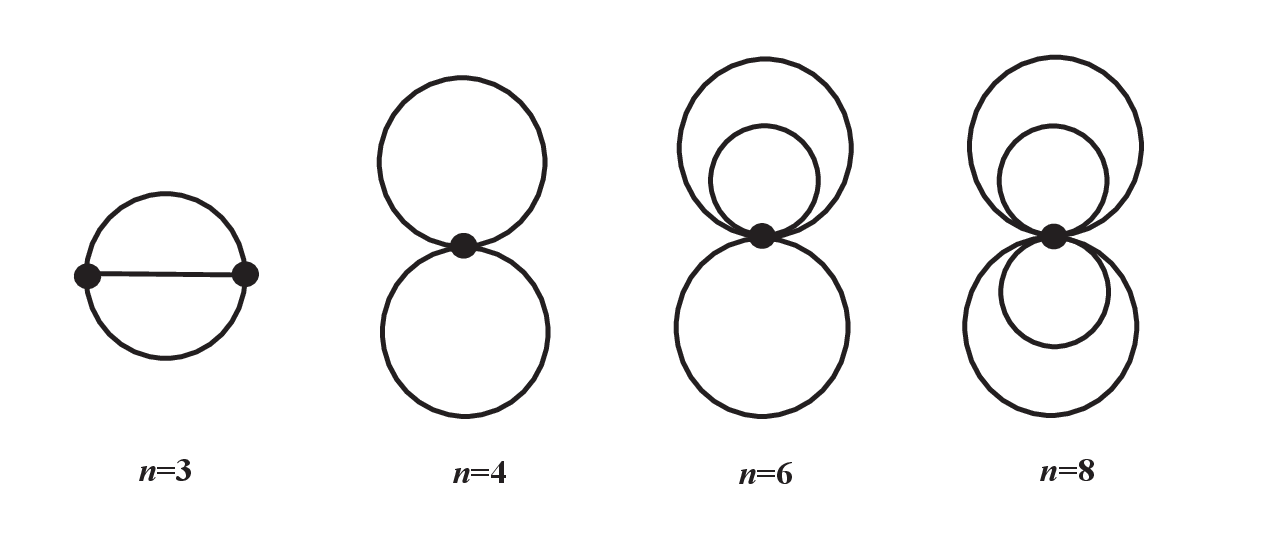}
	\caption{ The Feynman  bubble diagrams for $\lambda \Phi^n$ theory at the leading order. For $n=3$ one is dealing with a nested (double time) in-in integral   while for even values of $n=4, 6, 8...$  one has a single time in-in integral. For each even value of $n$, one has to consider a diagram with $\frac{n}{2}$ loops to calculate the order $\lambda$ corrections in expectation values such as $\langle \rho \rangle_\Omega$.
}
\label{fig}
\end{figure}

In this work we study the quantum fluctuations of a self-interacting  scalar field with non-minimal coupling to gravity in a dS background.
The free quantum fields with the conformal coupling in dS background \cite{Bunch:1978yq, Bunch:1978yw, Avis:1977yn, Ford:1977in,  Anderson:1985cw, Anderson:1987yt, Buchbinder:1987fg,Odintsov:1988wz,Odintsov:1990qq, Armendariz-Picon:2023gyl, Zhang:2019urk, Ye:2022tgs, Akhmedov:2022whm, Akhmedov:2020qxd, Akhmedov:2019esv}
and  the self-interacting scalar field with a quartic potential $V \propto \lambda \Phi^4$ \cite{ Onemli:2002hr, Onemli:2004mb, Brunier:2004sb, Miao:2005am, Prokopec:2008gw, Janssen:2008px,  Kahya:2009sz, Miao:2010vs, Glavan:2021adm, Glavan:2020gal, Woodard:2023vcw, Beneke:2023wmt}  are vastly studied in the literature. In this paper we extend those works to more general self-interacting potentials  with the emphasize on vacuum zero point energy associated to the bubble diagrams.  More specifically, we study potentials of the form $V \propto \lambda \Phi^n$ for even values of $n=4, 6, ...$ and calculate the expectation values of the
vacuum zero point energy and pressure associated to the bubble diagrams.
We employ dimensional regularization scheme \cite{tHooft:1972tcz, tHooft:1974toh, Bollini:1972ui,  Deser:1974cz, Dowker:1975tf, Barvinsky:1985an}
to regularize  the quantum infinities (for earlier works concerning dimensional  regularization scheme in dS spacetime see for example   \cite{ Onemli:2002hr, Onemli:2004mb, Brunier:2004sb, Miao:2005am, Prokopec:2008gw, Janssen:2008px,  Kahya:2009sz, Miao:2010vs, Glavan:2021adm, Glavan:2020gal, Woodard:2023vcw, Beneke:2023wmt}).  This paper extends  our earlier work  \cite{Firouzjahi:2023wbe} in which the vacuum zero point energy  associated to  the bubble diagram for a free field with a conformal coupling in a dS background   were studied. In the presence of  self-interaction, we have to consider multiple bubble loop diagrams as depicted in Fig. \ref{fig}. For an even value of $n$, we have to consider a bubble diagram with $\frac{n}{2}$ loops to obtain the order $\lambda$ corrections in the expectation values of physical quantities.

\section{The Setup}
\label{Mathematics}

We consider a real scalar field $\Phi$ in a dS spacetime which is non-minimally coupled to gravity with the conformal coupling $\xi$ and the
self-interacting potential $V(\Phi) = \lambda \Phi^n/n$ in which $\lambda$ is the self-interacting constant. As will become evident soon, we consider even values of $n$ with $n=4, 6, ...\, $. For the special case $n=4$, the coupling constant $\lambda$ is dimensionless while for higher values of $n$ it has the dimension
$M^{4-n}$. In addition, as we are interested in the conformal limit, and also to simplify the analysis, we assume the field is massless, $m=0$. However, as we shall see below, one can easily restore the mass in our formalism, though the equations will become more complicated.  In our analysis below, we treat the contribution of the self-interaction as a perturbation to the free theory and calculate the vacuum zero point energy and pressure to first order in coupling constant $\lambda$.

With the above discussions in mind, the action involving the scalar field  is given by
\begin{equation}
\label{Action00}
S=\int d^D x \sqrt{-g_{_D}}\left({-\frac{1}{2}\xi\Phi^2R} -\frac{1}{2} \nabla^\mu \Phi \nabla_ \mu \Phi -{\frac{\lambda}{n}\Phi^n}\right)\,,
\end{equation}
where $D$ refers to the dimension of the spacetime and $g_{_D}$ stands for the determinant of the metric.  Since we employ dimensional regularization to regularize the quantum infinities, we keep the spacetime dimension general and  set $D=4-\epsilon$ with $\epsilon\rightarrow 0$ as in conventional dimensional regularization approach. We work in the test field limit where the background geometry is the solution of the  Einstein field equations  with no backreactions from the scalar fields. In order for this approximation to be consistent, the vacuum zero point energy and pressure associated with the fluctuations of $\Phi$ should be much smaller than the corresponding background quantities.

In the absence of the self-interaction  the theory is classically conformal invariant in four dimensional spacetime when $\xi =\frac{1}{6}$. However, as it is well-known,  this classical symmetry is anomalous under quantum corrections. Furthermore, the addition of the potential can break the conformal invariance at the classical level since the
coupling constant $\lambda$ induces scale into the theory when $n \neq 4$. While the theory is still classically conformal invariant for $n=4$, but the quantum corrections break conformal invariance in this case as well. We consider the arbitrary values of even $n$ while  the special case of $n=4$ was studied extensively in the literature, see for example the works of Woodard and collaborators \cite{ Onemli:2002hr, Onemli:2004mb, Brunier:2004sb, Miao:2005am, Prokopec:2008gw, Janssen:2008px,  Kahya:2009sz, Miao:2010vs, Glavan:2021adm, Glavan:2020gal, Woodard:2023vcw, Beneke:2023wmt}.


The background spacetime has the form of the FLRW metric,
\begin{equation}\label{metric}
ds^2 = a(\tau)^2 \big( -d \tau^2 + d {\bf x}^2 \big) \, ,
\end{equation}
where  $\tau$ is the conformal time related to cosmic time via $dt = a(\tau) d \tau $ in which $a(\tau)$ is the scale factor. In our limit of a fixed dS background,
$a H \tau =-1 $,  in which $H$ is the Hubble expansion rate associated to the dS background. Since the dS spacetime is maximally symmetric, the Ricci tensor and Ricci scalar are given as follows,
\ba
\label{maximal}
R_{\mu \nu} = (D-1)H^2 g_{\mu \nu} , \quad \quad R= D (D-1) H^2 \, .
\ea

The scalar field equation   is given by
\begin{equation}
\label{FieldEQ}
{\Box\Phi  - \xi R\Phi  -{\lambda}{\Phi}^{n -1} = 0}\,.
\end{equation}
Note that to simplify the above field equation, we use the convention that the coupling constant has the form $\lambda/n$ instead of  $\lambda/n!$ which is usually used in QFT textbooks.

To study  the quantum fluctuations, we  introduce the canonically normalized  field $\sigma(\tau)$
\ba
\mathrm{\sigma(\tau)}\equiv a^{\frac{D-2}{2}}\Phi(\tau) \, ,
\ea
in terms of which the action takes the following diagonal form,
\ba
\label{Action01}
S=\frac{1}{2} \int d \tau d^{D-1} {\bf x}\left[\sigma^{\prime}(\tau)^{2}-(\nabla\sigma)^2+\left(\frac{(D-4)(D-2)}{4}\big(\frac{a^{\prime}}{a}\big)^2+\frac{D-2}{2} \frac{a^{\prime \prime}}{a}-({\xi R}+{\frac{\lambda}{a^D}\sigma^2}) a^2\right)\sigma^2\right]
\ea
where a prime indicates the derivative with respect to the conformal time.

To quantize the field  we expand it in terms of the creation and annihilation operators  in $D-1$ dimension Fourier space as follows,
\begin{equation}
\label{quantized-sigma}
\sigma\left(x^\mu\right)=\int \frac{d^{D-1} \mathbf{k}}{(2 \pi)^{\frac{(D-1)}{ 2}}}\left(\sigma_k(\tau) e^{i \mathbf{k}\cdot\mathbf{x}} a_{\mathbf{k}}+\sigma^*_k(\tau) e^{-i \mathbf{k} \cdot \mathbf{x}} a_{\mathbf{k}}^{\dagger}\right)\,,
\end{equation}
in which $\sigma_k(\tau)$ is the quantum mode  function and
$a_{\mathbf{k}}$ and $a_{\mathbf{k}}^{\dagger}$ satisfy the following
commutation relation in $D-1$ spatial dimension,
\begin{equation}
\label{Noncommutative}
\left[a_{\mathbf{k}}, a_{\mathbf{k^\prime}}^{\dagger}\right]=\delta^{D-1}(\mathbf{k}-\mathbf{k^\prime})\,.
\end{equation}
Correspondingly,  the equation of motion of the {free} mode function (with $\lambda=0$)  from the action \eqref{Action01} is given by
\begin{equation}
\label{EquationofMotion}
\sigma_k^{\prime \prime}(\tau)+\left[k^2+ \frac{1}{\tau^2} \Big( {{D (D-1)}} \xi-\frac{D(D-2)}{4} \Big) \right] \sigma_k(\tau)=0\,.
\end{equation}
Note that the above equation is similar to the Mukhanov-Sasaki equation associated to the  inflaton perturbations in an inflationary background \cite{Mukhanov:1990me}.
 Notice that for  $\xi =\frac{1}{6}$ in $D=4$ the second term in the big bracket vanishes and the mode function reduces to its simple flat form. This corresponds to the conformal limit which we consider in our analysis below. However, note that  in a general $D$-dimensional spacetime, the conformal limit corresponds to the special value
\ba
\xi= \xi_D\equiv \frac{D-2}{4 (D-1)} \, .
\ea

Imposing the Bunch-Davies (Minkowski) initial conditions for the modes  deep inside the horizon, the
solution of the mode function from Eq. (\ref{EquationofMotion})
is given in terms of the Hankel function,
\begin{equation}\label{Phi-k}
\Phi _k(\tau) = a^{\frac{{2 - D}}{2}}{\sigma _k}(\tau ) = {( - H\tau )^{\frac{{D - 1}}{2}}}{\left( {\frac{\pi }{{4H}}} \right)^{\frac{1}{2}}} {{e^{\frac{i \pi}{2}  (\nu + \frac{1}{2})} }}
H_\nu ^{(1)}( - k\tau ){\mkern 1mu}\,,
\end{equation}
where
\begin{equation}
\label{nu-00}
\nu  \equiv  \frac{1}{2}{\mkern 1mu} \sqrt {(D-1)^2 {-4 D (D-1) \xi}} \, .
\end{equation}
From the above expression  we see that $\nu$ can be either real or pure imaginary depending on the values of $\xi$.  In our limit of interest where
$\xi$ is near its conformal value, $\xi$ is real. In particular, for $\xi= \frac{1}{6}$ with $D=4$, we obtain $\nu =\frac{1}{2}$. As the in-in integrals become non-trivial, and in order to prevent complications associated to an imaginary $\nu$ in the mode functions, in our analysis below we assume that $\nu$ is real. This imposes the bound $0<\xi \leq \frac{D-1}{4 D}$ which for $D=4$ corresponds to $\xi\leq \frac{3}{16}$.


The energy momentum tensor in the presence of self-interaction is given by,
\begin{eqnarray}\label{E-Mtensor-nonmimimal}\nonumber
{{T_{\mu \nu }}}& =& (1 - 2\xi ){\partial _\mu }\Phi {\partial _\nu }\Phi  + (2\xi  - \frac{1}{2}){g_{\mu \nu }}{g^{\alpha \beta }}{\partial _\alpha }\Phi {\partial _\beta }\Phi \\
\,\,\,\,\,\,\,\, &+ &\xi ({R_{\mu \nu }} - \frac{1}{2}{g_{\mu \nu }}R){\Phi ^2} + 2\xi ({g_{\mu \nu }}\Phi\Box \Phi  - \Phi {\nabla _\nu }{\nabla _\mu }\Phi )  {-\frac{\lambda}{n}{g_{\mu \nu }}{\Phi ^n}}\, .
\end{eqnarray}
Employing the field equation (\ref{FieldEQ}) one can  eliminate $\Box \Phi$ and
using  Eq. (\ref{maximal}),  $T_{\mu \nu}$ is further simplified to
\begin{eqnarray}
\label{T-simple}
\begin{array}{l}
{T_{\mu \nu }} = {\partial _\mu }\Phi {\partial _\nu }\Phi  + \frac{{{g_{\mu \nu }}}}{2}(4\xi  - 1){\partial ^\alpha }\Phi {\partial _\alpha }\Phi  \\
\,\,\,\,\,\,\,\,\,\, + \frac{\xi }{2}(D - 1)\Big( 2 + (4\xi  - 1)D \Big){H^2}{g_{\mu \nu }}{\Phi ^2} - \xi {\nabla _\mu }{\nabla _\nu }{\Phi ^2} + {{g_{\mu \nu }}{\lambda }(2\xi  - \frac{1}{n}){\Phi ^n}}\,.
\end{array}
\end{eqnarray}
Similarly, the trace of the energy momentum-tensor  $T\equiv T^\mu_\mu$ is  obtained to be
\begin{eqnarray}\label{T-trace00}
\begin{array}{l}
\label{T-eq}
T = 2 \Big( (D-1) \xi + \frac{2-D}{4} \Big) \Big( \partial^\alpha \Phi \partial_\alpha \Phi + D (D-1) \xi H^2 \Phi^2  \Big) + {{\lambda }\Big(2\xi (D-1) - \frac{D}{n}\Big){\Phi ^n}}\, .
\end{array}
\end{eqnarray}
The  energy density  $\rho = T_{00}$ is:
\ba
\label{rho-eq0}
 \rho   &=& \frac{(1+ 4 \xi)}{2}  \dot \Phi^2  +
\frac{(1- 4 \xi)}{2}     \nabla^i \Phi \nabla_i \Phi  +
\frac{H^2}{2} \Big[  (1- 4 \xi)  {{\big(  D (D-1) \xi)}} - 2 (D-1) \xi  \Big]  \Phi^2
\nonumber\\
&-& \xi \nabla_0 \nabla_0  \Phi^2 - \lambda ( 2 \xi - \frac{1}{n} ) \Phi^n \, .
\ea

Finally,  the pressure $P$ is given by
\ba\label{Pressure}
P = \frac{1}{{D - 1}}{ \bot ^{\mu \nu }}{T_{\mu \nu }} \, ,
\ea
in which $\perp^{\mu \nu} \equiv g^{\mu \nu}+u^\mu u^\nu$ represents the  projection operator and $u^\mu=(1,0,0,0)$ is the comoving four velocity. Consequently, we obtain
\ba
\label{P-eq}
P= \frac{1}{ {D - 1} }( T+ \rho) \, .
\ea

In our analysis below, we will be mainly interested in vacuum expectation values
such as $\langle \rho \rangle $, $\langle P \rangle $ and $\langle T \rangle $. This was studied for a free theory with a non-zero mass in \cite{Firouzjahi:2023wbe} and here we extend these analysis in the presence of the
self-interaction $\lambda \Phi^n$. However, it is important to note that the expectation value is with respect to the full vacuum which we denote
by $| \Omega \rangle$ so $\langle \rho \rangle \equiv \langle \Omega | \rho| \Omega \rangle$ and so on.  Because of the interaction term $\lambda \Phi^n$ the
vacuum  $| \Omega \rangle$  is different  than the vacuum associated to the free theory which is denoted by $| 0 \rangle$. To prevent confusion we define $\langle \Omega | \rho| \Omega \rangle \equiv \langle \rho \rangle_{\Omega} $ while  $\langle 0 | \rho| 0 \rangle \equiv \langle \rho \rangle_0 $ and so on.


\section{Dimensional Regularizations and In-In Formalism }
\label{dim-reg}

In this section we calculate vacuum expectation values such as $\langle \rho \rangle_\Omega$  using dimensional regularization scheme in $D$ dimension. As in \cite{Firouzjahi:2023wbe}, let us define
\ba
\label{rhoi}
\rho_1 \equiv \frac{1}{2} \dot \Phi^2 \, , \quad \quad
\rho_2 \equiv \frac{1}{2} g^{i j} \nabla_i \Phi \nabla_j \Phi \, , \quad \quad
\rho_3 \equiv \frac{1}{2} H^2 \Phi^2 \, .
\ea
Then,
\ba
\label{rho-av}
\langle \rho \rangle_\Omega &=& ( 1 + 4 \xi ) \langle \rho_1\rangle_\Omega
+  ( 1 - 4 \xi ) \langle \rho_2\rangle_\Omega  + \Big[ ( 1- 4 \xi)  D (D-1) \xi - 2 (D-1) \xi \Big] \langle \rho_3\rangle_\Omega \nonumber\\
&-& \xi \langle \nabla_0 \nabla_0 \Phi^2 \rangle_\Omega -\lambda  (2 \xi -\frac{1}{n}) \langle \Phi^n \rangle_\Omega \, .
\ea
Note that the first three terms in the first line in Eq. (\ref{rho-av}) are formally the same as in \cite{Firouzjahi:2023wbe} except that in \cite{Firouzjahi:2023wbe} the expectation values were with respect to the vacuum of the free theory. The two terms in second line in Eq. (\ref{rho-av}) are new. First, we have a term with
the specific $\lambda$ coupling. Second, the contribution $\langle \nabla_0 \nabla_0 \Phi^2 \rangle_\Omega$ which is non-trivial. In the analysis of
\cite{Firouzjahi:2023wbe} it was noticed that $\langle \nabla_0 \nabla_0 \Phi^2 \rangle_0 =0$. This is because $\langle \Phi^2 \rangle_0$ is a constant so it is easy to understand  that $\langle \nabla_0 \nabla_0 \Phi^2 \rangle_0 =
\nabla_0 \nabla_0 \langle \Phi^2 \rangle_0 =0$. However, in the presence of the interaction, we notice that $\nabla_0 |\Omega \rangle \neq 0$ so one can not simply take the derivative outside the expectation value, i.e.
$\langle \nabla_0 \nabla_0 \Phi^2 \rangle_\Omega \neq
\nabla_0 \nabla_0 \langle \Phi^2 \rangle_\Omega $.

Out of the five contributions into $\langle \rho \rangle_\Omega$ in Eq. (\ref{rho-av}), the last term  $\lambda \langle \Phi^n \rangle_\Omega$ is the easiest term to calculate. This is because it has a factor of $\lambda$ and since we are interested in first order corrections in $\lambda$, we can simply assume the vacuum in this case is the free vacuum   and
\ba
\lambda   \langle \Phi^n \rangle_\Omega \simeq
\lambda   \langle \Phi^n \rangle_0 \,  + {\cal O}(\lambda^2) \, .
\ea
Note that the assumption that $n$ is even was necessary to obtain the above result to leading order in $\lambda$. For odd values of $n$, the linear term in $\lambda$ vanishes and one has to go to higher orders of $\lambda$ to calculate
$\lambda   \langle \Phi^n \rangle_\Omega $. This brings additional complexities involving in-in integral as we shall see in next section.

Since $\Phi$ is a Gaussian free field in the absence of interaction, one can easily see that for even values of $n$
\ba
  \langle \Phi^n \rangle_0 \simeq  (n-1)!!
  \big( \langle \Phi^2  \rangle_0 \big)^\frac{n}{2} \, ,
  \ea
in which $(n-1)!! = (n-1) (n-3)...1\, $. For example, for $n=4$, we have
$\langle \Phi^4 \rangle_0 = 3 \big( \langle \Phi^2 \rangle_0 \big)^2$.
As a result, the last term in Eq. (\ref{rho-av}) reads
\ba
\label{last-rho}
\lambda  (2 \xi -\frac{1}{n}) \langle \Phi^n \rangle_\Omega \simeq
 (2 \xi -\frac{1}{n})  \lambda \big( \langle \Phi^2  \rangle_0 \big)^\frac{n}{2}
 + {\cal O}(\lambda^2) \, .
\ea
The quantity $ \langle \Phi^2  \rangle_0 $ can be viewed as the coincident limit of the Feynman propagator. It plays important roles in our analysis below in which the  expectation values of the physical quantities
in the presence of interaction can be expressed in terms
of $ \langle \Phi^2  \rangle_0 $.

\subsection{Free Theory}

Here we briefly review the results of \cite{Firouzjahi:2023wbe} for the free theory which will be used in our following analysis as well. In the free theory with $\lambda=0$, the vacuum $| 0 \rangle$ is dS invariant so from Eq. (\ref{rho-av}) we obtain
\ba
\label{rho-av-vac}
\langle \rho \rangle_0 = ( 1 + 4 \xi ) \langle \rho_1\rangle_0
+  ( 1 - 4 \xi ) \langle \rho_2\rangle_0  + \Big[ ( 1- 4 \xi)  D (D-1) \xi - 2 (D-1) \xi \Big] \langle \rho_3\rangle_0 \, .
\ea
As we shall see below, all three components of $ \langle \rho_i  \rangle_0 $
are expressed in terms of $ \langle \Phi^2  \rangle_0 $
so let us calculate this quantity. Performing the dimensional regularization analysis, we have
\ba
\langle \Phi^2  \rangle_0  ={ \mu ^{4 - D} }
\int \frac{ d^{D - 1} {\bf k}}{(2\pi )^{D - 1} }  \left| \Phi _k (\tau ) \right|^2 \, ,
\ea
in which $\mu$ is a mass scale to keep track of the dimensionality of physical quantities. We decompose the integral into the radial and angular parts as follows
\ba
{{\rm{d}}^{D-1}}{\bfk} = {k^{D - 2}}\;{\rm{d}}k\; {{{\rm{d}^{D-2}} }}\Omega {\mkern 1mu}\, ,
\ea
where  ${\rm{d}^{D-2}}\Omega$ represents the $D-2$-dimensional  angular part
with the volume
\ba
\label{D-2-vol}
 \int \mathrm{d}^{D-2} \Omega=\frac{ 2 \,  \pi^{\frac{D-1}{2}}}{\Gamma\left(\frac{D-1}{2}\right)}\,.
\ea
Combining all numerical factors and defining the dimensionless variable $x\equiv - k \tau$ we obtain
\ba
\label{rho3}
\langle \Phi^2  \rangle_0=
{\frac {{\pi}^{\frac{3-D}{2}  }{\mu}^{4-D}{H}^{D-2}}{{2}^{D}\Gamma \left( \frac{D-1}{2} \right) }}   \int_0^{\infty} dx~x^{D-2}  \left| H_{ \nu}^{(1)}(x)\right|^2 \, .
\ea
Performing the integral, this yields  \footnote{We use the Maple computational software to calculate the integrals. }
\begin{eqnarray}\label{Phi2}
\langle \Phi^2  \rangle_0= \frac{{{\mu ^{4 - D}}{\pi ^{ - \frac{D}{2} - 1}}}}{2^D} \Gamma \Big( {\nu  + \frac{D}{2} - \frac{1}{2}} \Big)\Gamma \Big( { - \nu  + \frac{D}{2} - \frac{1}{2}} \Big)\Gamma \Big( { - \frac{D}{2} + 1} \Big)\cos \big( {\pi {\mkern 1mu} \nu } \big){ H^{D-2} }\,.
\end{eqnarray}
In particular, note that for the conformal limit with $\nu= \frac{1}{2}$, the above expression vanishes. With  $\langle \Phi^2  \rangle_0$ at hand and using Eq. (\ref{last-rho}), the last term for $\langle \rho\rangle_\Omega$
in Eq. (\ref{rho-av}) is calculated accordingly.

Following   similar steps to calculate  $\langle \rho_i \rangle_0$  we obtain the following relations \cite{Firouzjahi:2023wbe}
\ba
\label{rho13}
\langle \rho_1 \rangle_0 =  (D-1) \xi  \langle \rho_3 \rangle_0 \, , \quad  \langle \rho_2 \rangle_0 = -(D-1) \langle \rho_1 \rangle_0 =
- (D-1)^2 \xi  \langle \rho_3 \rangle_0\,,
\ea
where from Eq. \eqref{rhoi}, $ \langle \rho_3 \rangle_0 = \frac{H^2}{2} \langle \Phi^2  \rangle_0 $ with $\langle \Phi^2  \rangle_0$ given in Eq. (\ref{Phi2}).   In the conformal limit where $\langle \Phi^2  \rangle_0=0$, we see that
$\langle \rho_i \rangle_0=0$ and correspondingly $\langle \rho \rangle_0=0$.

Similarly, calculating $\langle T \rangle_0 $ one can show that
$\langle T \rangle_0= -D \langle \rho \rangle_0$, and correspondingly \cite{Firouzjahi:2023wbe}
 \ba
 \label{rho-P}
 \langle P \rangle_0 = - \langle \rho \rangle_0 \, .
 \ea
This is the expected result indicating the local Lorentz invariance for the free theory in which one expects to locally have $\langle T_{\mu \nu} \rangle = -\langle \rho \rangle g_{\mu \nu}$.

It is important to note that in the above results, $D$ is general so to perform the regularization we consider $D= 4-\epsilon$. One sets $\epsilon \rightarrow 0$ at the end  with the understanding that the singular terms in physical quantities
with inverse powers of $\epsilon$ are cancelled by  appropriate counter terms as in standard QFT analysis.

It is useful to look at the results in some limits of interests.
In the case of massless field with no conformal limit, $\xi=0$, one obtains
\ba
\label{massless-rho}
\langle \rho \rangle_0^{\mathrm{reg}}= \frac{3 H^4}{32 \pi^2}
= -\frac{1}{4} \langle T \rangle_0^{\mathrm{reg}}\, ,
 \quad \quad \quad  (\xi=0)  \, .
\ea
On the other hand, as we noticed before, for the special case of conformal point with  $ \xi= \frac{1}{6}$,
$\langle \rho \rangle_0^{\mathrm{reg}}= \langle T \rangle_0^{\mathrm{reg}}=0$.
However, if one restores the mass so the theory is not conformally invariant,  one obtains \cite{Firouzjahi:2023wbe}
\ba
\langle \rho \rangle_0^{\mathrm{reg}}= -\frac{ H^4}{96 \pi^2} \beta^2
 + \frac{ H^4}{64 \pi^2} \Big[ \ln\Big( \frac{H^2}{4\pi \mu^2 }  \Big)  - \frac{1}{2} \Big] \beta^4 +  {\cal O}( \beta^6) \, , \quad \quad \quad
(\xi=\frac{1}{6} ) \, ,
\ea
in which $\beta \equiv m/H$.

\subsection{In-in formalism}

To calculate the first four terms in Eq. (\ref{rho-av}) to first order in $\lambda$, we need to implement the in-in formalism which perturbatively relates the vacuum expectation values of the interacting theory to the vacuum expectation of the free theory as follows \cite{Weinberg:2005vy}
\begin{equation}
\label{in-in definition}
\langle {\cal O}(\tau) \rangle_\Omega= \Big\langle 0 \Big| \bar{T} e^{i \int_{\tau_0}^\tau H_I\left(\tau^{\prime}\right) d \tau^{\prime}} {\cal O}_I(\tau)T e^{-i \int_{\tau_0}^\tau H_I\left(\tau^{\prime}\right) d \tau^{\prime}} \Big|0  \Big\rangle \,,
\end{equation}
where $T$ and $\bar{T}$ stand for time ordering and anti-time ordering respectively. The subscript $I$ in the right hand side of the above equation indicates that all quantities are calculated in the interaction picture, i.e. with the mode function of the free theory given by Eq. (\ref{Phi-k}). The initial time
is $\tau_0=-\infty$ while the final slicing $\tau$ is the time when the measurement on the quantum operator ${\cal O}$ is made. As we work in an unperturbed dS
background, we have  $-\infty <\tau' \leq  \tau <0$. Since the time integrals in Eq. (\ref{in-in definition}) are non-trivial, we shall restrict ourselves to the case where the upper limit $\tau \rightarrow 0$, i.e. the measurement is being made towards the future boundary of dS. In an inflationary setup with deviations from an exact dS background, the final slicing $\tau \rightarrow 0$ corresponds to the time of end of inflation.
Finally,  $H_I$ represents  the interacting Hamiltonian, which in our case
 is
 \ba
 H_I = \frac{\lambda }{n} a^D \int d^{{D-1}} {\bf x} \,  \Phi(x)^n \, .
 \ea
 Note that the factor $a^D$ appears because of the volume element $\sqrt{-g}$.

 To leading order in $\lambda$, the correction in $\langle {\cal O}(\tau) \rangle_\Omega$ is given by
 \ba
 \label{first-correction}
 \langle {\cal O}(\tau) \rangle_\Omega= \langle {\cal O}(\tau) \rangle_0
 +2 \mathrm{Im} \int_{-\infty}^\tau d \tau' \big \langle  {\cal O}(\tau) H_I (\tau') \big\rangle_0 \, .
 \ea
The first term above is the contribution in the absence of interaction which were
calculated in \cite{Firouzjahi:2023wbe} for  ${\cal O}= \rho, P$ etc.  Our goal below is to calculate the above integrals for four different operators $\rho_1, \rho_2, \rho_3$ and $\nabla_0 \nabla_0 \Phi^2$ as appearing in Eq. (\ref{rho-av}). In order to isolate the contribution of the free theory, we denote the last term in Eq. (\ref{first-correction}) by ${\Delta}\langle {\cal O}(\tau) \rangle$.

Let us start with  $\langle {\rho_3} (\tau) \rangle_\Omega$, yielding
\ba
\label{rho3-int1}
{\Delta} \langle \rho_3(\tau, {\bf x_0}) \rangle &=&
\frac{H^2}{2} {\Delta} \langle \Phi^2(\tau, {\bf x_0}) \rangle \nonumber\\
&=&  \frac{\lambda H^2}{n} \Im\Big[  \int_{-\infty}^\tau d \tau' a(\tau')^D \int d^{{D-1}}  x
\big \langle \Phi(\tau, {\bf x_0})^2  \Phi(\tau' , {\bf x})^n  \big\rangle_0 \Big]  \, .
\ea
Note that ${\bf x_0}$ is an arbitrary reference point in background where the measurement is made. However, because of the spatial translation invariance, we
can set ${\bf x_0}=0$ so we do not specify  ${\bf x_0}$ in the rest of the analysis below. From the above expression we see that for odd values of $n$, the expectation values vanish in the light of Wick theorem. Therefore, for odd values of $n$ one needs to go to second order of perturbations, leading to order ${\cal O}(\lambda^2)$ corrections. This in turn requires
nested  time integrals (i.e. a double time integral)  which are more complicated than the single time integral over $\tau'$ which we encounter for even value of $n$ as given in Eq. (\ref{rho3-int1}). For this reason, as mentioned before, we restrict our analysis to even values of $n=4, 6, ...\, $.

There are two different types of contributions when performing the contractions
in Eq. (\ref{rho3-int1}). The first type is in the form
$\langle \Phi(\tau, {\bf x_0})^2 \rangle_0 \langle   \Phi(\tau' , {\bf x})^n\rangle_0 $. With a bit of efforts one can show that this contribution has no imaginary component so this contribution vanishes. The second type contracts each term of $\Phi(\tau, {\bf x_0})$ with a term in $\Phi(\tau' , {\bf x})^n$. There are total $n (n-1)!!$ possibilities for these contractions. After performing the Wick contractions we obtain (for further details of Wick contractions see Appendix \ref{contractions}),
\ba
\label{rho3-int2}
\Delta \langle \rho_3(\tau) \rangle= {(n-1)!!} \lambda H^2   \,
\big( \langle \Phi^2  \rangle_0 \big)^\frac{n-2}{2} \,   \mu^{4- D} I_3 (\tau)  \, ,
\ea
in which
\ba
\label{I3}
I_3 (\tau)  \equiv
 \int_{\tau_0}^\tau d \tau' a(\tau')^D
\int \frac{d^{D-1} \bfq }{(2 \pi)^{(D-1)}} \mathrm{Im} \Big[  \Phi_q( \tau)^2 {\Phi_q^*}( \tau')^2\Big]  \, .
\ea

Following similar steps for $\rho_1$ and $\rho_2$ we have
\ba
\label{rho1-int1}
\Delta \langle \rho_1(\tau) \rangle= {(n-1)!!} \lambda H^2
\big( \langle \Phi^2  \rangle_0 \big)^\frac{n-2}{2} \,   \mu^{4- D} I_1(\tau) \, ,
\ea
with
\ba
\label{I1}
I_1 (\tau)  \equiv
 \int_{-\infty}^\tau d \tau' {\frac{a(\tau')^D}{a(\tau)^2}}   \int \frac{d^{D-1} \bfq }{(2 \pi)^{(D-1)}} \mathrm{Im} \Big[  \Phi_q'( \tau)^2 {\Phi_q^*}( \tau')^2\Big] \, ,
\ea
and
\ba
\label{rho2-int1}
\Delta \langle \rho_2(\tau) \rangle=  {(n-1)!!} \lambda H^2
\big( \langle \Phi^2  \rangle_0 \big)^\frac{n-2}{2}   \mu^{4- D}  I_2(\tau)  \, ,
\ea
with
\ba
\label{I2}
I_2 (\tau)  \equiv
 \int_{-\infty}^\tau d \tau' {\frac{ q^2a(\tau')^D}{a(\tau)^2}} \int \frac{d^{D-1} \bfq }{(2 \pi)^{(D-1)}}  \mathrm{Im} \Big[  \Phi_q( \tau)^2 {\Phi_q^*}( \tau')^2\Big]  .
\ea
In addition, we have to calculate $\langle \nabla_0 \nabla_0 \Phi^2 \rangle_\Omega$ as well, which is given by
\ba
\langle \nabla_0 \nabla_0 \Phi^2 \rangle_\Omega = \frac{2}{a^2} \Big[ \langle \Phi'^2 \rangle_\Omega
+ \langle \Phi \Phi'' \rangle_\Omega + \frac{1}{\tau} \langle \Phi \Phi' \rangle_\Omega
\Big] \, .
\ea
Calculating each term as above, we obtain
\ba
\label{Phi-tt}
\langle \nabla_0 \nabla_0 \Phi^2 \rangle_\Omega = {(n-1)!!} \lambda H^2
\big( \langle \Phi^2  \rangle_0 \big)^\frac{n-2}{2}  \mu^{4- D}  I_4(\tau) \, ,
\ea
in which
\ba
\label{I4}
I_4 (\tau)  \equiv
  \int_{-\infty}^\tau d \tau' {\frac{ a(\tau')^D}{a(\tau)^2}}   \int \frac{d^{D-1} \bfq }{(2 \pi)^{(D-1)}}   \mathrm{Im} \Big[ \Big( \Phi_q'( \tau)^2  + \Phi_q( \tau) \Phi''_q( \tau)  +\frac{1}{\tau} \Phi_q( \tau) \Phi'_q( \tau) \Big)  {\Phi_q^*}( \tau')^2\Big] .
\ea

\section{Measurements at Future dS Boundary}

So far our analysis were general except that we assumed that $n$ is even so we deal with a single time integral  as in Eq. (\ref{I3}). To proceed further, we should calculate each of $I_i(\tau)$ listed above. We start with $I_3(\tau)$ which is easier.  Let us denote the $(D-2)$ dimensional angular part of the momentum integral
by $ V_{D-2}$ as given in Eq. (\ref{D-2-vol}). Defining the dimensionless variables $x\equiv - q \tau'$ and $y\equiv - q \tau$ and switching the orders of the time and  momentum integrals, we obtain
\ba
I_3= \frac{V_{D-2} }{(2 \pi)^{D-1}}  \Im\Big[ \frac{\pi^2 H^{D-4} }{16 } \int_0^\infty d y  y^{D-2} \big(H_\nu^{(1)} (y) \big)^2   \int_y^\infty \frac{dx}{x}  \big( H_\nu^{(2)} (x) \big)^2  \Big]  \, .
\ea
Looking at the  integral over the $x$ variable we see that it is in the form of a nested integral. Furthermore, its integrand falls off quickly for large $x$ as the integrand  oscillates rapidly with a decaying amplitude. Therefore, one expects the dominant contribution for the interior integral comes from the lower bound when $x \rightarrow y$.

To proceed further and to calculate the integrals analytically, we have to impose some simplification approximations. As a reasonable approximation,  we take $\tau \rightarrow 0$. This corresponds to performing the measurement at the future boundary of dS. In inflationary models, this corresponds to performing the vacuum expectation value at the time of end of inflation. In the limit $\tau \rightarrow 0$, the mode function $\Phi_q(\tau)$ in Eq. (\ref{Phi-k}) simplifies to
\ba
\label{phi-k-app}
\Phi_q(\tau) \rightarrow -\frac{i \Gamma(\nu)}{\pi}  \big( \frac{\pi}{ 4 H} \big)^{\frac{1}{2}} e^{ \frac{i \pi}{2} (\nu + \frac{1}{2}) } \big(-H \tau \big)^{\frac{{D}-1}{2}}
\big(  \frac{-2}{k \tau} \big)^{\nu} \, .
\ea
In the context of inflation, this represents the superhorizon limit of cosmological perturbations when $q \ll a H$ so $y=-q \tau \rightarrow 0$.

Expanding the interior integrand for $x \ll 1$ and taking the lower bound of integral with $x  \rightarrow y \rightarrow 0$, the rest of integral over $y$ can be taken analytically
yielding to
\ba
\label{I3-val}
I_3 = \frac{4^\nu {\pi}^{\frac{1}{2}} H^{D-4}  }{32 \nu} \frac{ \Gamma( - \nu +\frac{d}{2} - \frac{1}{2} ) \Gamma(  \nu -\frac{d}{2} +1)  \Gamma( \frac{d}{2} - \frac{1}{2} )}{  \Gamma( 2 \nu -\frac{d}{2} + \frac{3}{2}) \Gamma(1- \nu)^2  } \sin(\pi \nu) \sin\big(  \pi \nu - \frac{\pi D}{2}\big)   \, .
\ea

Following the same strategy, we can calculate $I_1, I_2$ and $I_4$ analytically. It turns out that $I_i$ are related to each other so all of them can be expressed in terms of $I_3$ as follows:
\ba
I_1= \frac{(4 \nu^2 + d- 2 \nu -1) (- 2 \nu + d -1)}{4 ( d- 2 \nu)} I_3 \, ,
\ea
\ba
I_2= -\frac{(d-1-4 \nu) (d-1 - 2 \nu) (d-1)}{4 (d - 2 \nu)} I_3
\ea
and
\ba
I_4= 8 \nu^2 I_3 \, .
\ea

With the above analytical values of the in-in integrals at hand, we can calculate
$\langle \rho \rangle_\Omega,  \langle P \rangle_\Omega$ and  $\langle T \rangle_\Omega$. The expressions for these quantities for a general value of $\xi$ are too complicated to report here so we consider two spacial limits for analytical presentations.  First, the conformal point where $\xi =\frac{1}{6}$ and second, the limit of small deviation from the conformal point with $\delta \xi \equiv \xi -\frac{1}{6} \ll 1$. For general values of
$\xi$ where the analytical results are intractable,
we present the numerical plots of $\langle \rho \rangle_\Omega$ and  $\langle T \rangle_\Omega$.

\subsection{Conformal point: $\xi= \frac{1}{6}$ }

At the conformal point with $\xi= \frac{1}{6}$, we obtain
\ba
\Delta \langle \rho \rangle =  \frac{\lambda {(3-n) } }{3 n}  (n-1)!! \Big(\frac{-H^2}{24 \pi^2} \Big)^\frac{n}{2} \, ,
\ea
and
\ba
\Delta \langle P \rangle =  \frac{\lambda (2n-9)  }{9 n}  (n-1)!! \Big(\frac{-H^2}{24 \pi^2} \Big)^\frac{n}{2} \, .
\ea


From the above formulas for $\Delta \langle \rho \rangle$ and $\Delta \langle P \rangle$, and noting that when $\lambda=0$ both $ \langle \rho \rangle$ and
$ \langle P \rangle$  vanish,   the equation of state $w= \frac{P}{\rho}$ is obtained to be
\ba
\label{w-eq}
w=  \frac{-2n + 9}{3n -9} \, .
\ea
For example for $n=4$, corresponding to two-loop quantum corrections, we obtain $w= \frac{1}{3}$ so the quantum corrections in energy momentum tensor behave like radiation. On the other hand, for large values of $n$, the equation of state approaches to $w \rightarrow -\frac{2}{3}$. It is intriguing that the quantum corrections from self-interactions are not in the form of  $w=-1$ which is expected from  local Lorentz invariance.

It is instructive to calculate
$\Delta \langle T \rangle$ as well, yielding
\ba
\Delta \langle T \rangle =  \frac{\lambda (n-4)  }{ n}  (n-1)!! \Big(\frac{-H^2}{24 \pi^2} \Big)^\frac{n}{2} \, .
\ea
Interestingly, for the case $n=4$, we see that the quantum corrections in the trace of energy momentum tensor vanish. This is consistent with the fact that for $n=4$ the parameter $\lambda$ is dimensionless so the theory is classically scale invariant and $T=0$ at the classical level.  It is intriguing  that the two-loop quantum correction respects this result as well. However, it is an open question whether or not higher orders loops corrections  (i.e. $\lambda^2$ and higher orders corrections)  respect this conclusion.

\subsection{Small deviation from conformal point }

Now suppose we slightly deviate from the conformal point with
$\delta \xi =\xi -\frac{1}{6} \ll 1$. We calculate the quantum corrections to leading order in $\delta \xi$. By increasing the value of $n$, the analysis become  complicated, so here we present the results for two cases $n=4$ and $n=6$.

Starting with $n=4 $ to linear order in $ \big( \xi -\frac{1}{6}  \big)$ we obtain
\ba\label{Deltarho-n4}
\Delta \langle \rho \rangle \simeq  -\frac{\lambda H^4}{2304 \pi^4}+
 \frac{\lambda H^4}{32 \pi^4} \Big[-\frac{1}{\epsilon}+  \ln\big( \frac{H^2}{4 \pi\mu^2}  \big) + \frac{5}{2} - \gamma \Big]  \big( \xi -\frac{1}{6}  \big)  + {\cal O} \big( \big( \xi -\frac{1}{6}  \big)^2 \big)\, ,  \quad (n=4)
\ea
and
\ba
\Delta \langle P \rangle \simeq   -\frac{\lambda H^4}{6912 \pi^4}+
 \frac{\lambda H^4}{96 \pi^4} \Big[ -\frac{1}{\epsilon}+ \ln\big( \frac{H^2}{4 \pi\mu^2}  \big) + \frac{23}{6} - \gamma \Big]  \big( \xi -\frac{1}{6})  + {\cal O} \big( \big( \xi -\frac{1}{6}  \big)^2 \big)\, , \quad (n=4) \, ,
\ea
in which $\gamma$ is the Euler number.
We have the divergent $\epsilon^{-1}$ term  in both
$\Delta \langle \rho \rangle $ and  $\Delta \langle P \rangle $ which appears when $\xi \ne  \frac{1}{6}$ which should be removed by  appropriate counter terms.
Interestingly, in this case we see that the singular terms in $\Delta \langle \rho \rangle $ and  $\Delta \langle P \rangle $
have the equation of state
associated to radiation, $w= \frac{1}{3}$, as suggested in Eq. (\ref{w-eq}) while this does not hold for the finite terms. Indeed, we have
\ba
\Delta \langle \rho \rangle - 3 \Delta \langle P \rangle \simeq  \frac{19 \lambda H^4}{96 \pi^4}  \big( \xi -\frac{1}{6}  \big) \quad  \quad \quad (n=4) \, ,
\ea
so the divergent  $\epsilon^{-1}$ terms are cancelled in the above expression.
In addition, the trace of energy momentum tensor is not zero,
\ba\label{DeltaTrace-n4}
\Delta \langle T \rangle \simeq   -\frac{3\lambda H^4}{16 \pi^4}  \big( \xi -\frac{1}{6}  \big)    \quad  \quad \quad (n=4) \, .
\ea
However, we notice that it has no singular part.

Now consider the case $n=6$, corresponding to three-loop bubble diagrams.  In this case the coupling constant $\lambda$ has the dimension of $M^{-2}$ so the quantum corrections would scale like $\lambda H^6$. To linear order in $ \big( \xi -\frac{1}{6}  \big)$ one obtains
\ba
\Delta \langle \rho \rangle \simeq  \frac{5 \lambda H^6}{27648 \pi^6}-
 \frac{15\lambda H^6}{512 \pi^6} \Big[-\frac{2}{3 \epsilon}  + \ln\big( \frac{H^2}{4 \pi\mu^2}  \big) + \frac{71}{54} - \gamma \Big]  \big( \xi -\frac{1}{6}  \big)
 + {\cal O} \big( \big( \xi -\frac{1}{6}  \big)^2 \big) ,   \quad (n=6) .
\ea
As expected, we have the singular term $\epsilon^{-1}$. In addition, there will be singular terms $\epsilon^{-2}$ but it comes at second order  $ \big( \xi -\frac{1}{6}  \big)^2$.
Similarly, for the pressure we obtain
\ba
\Delta \langle P \rangle \simeq \frac{-5 \lambda H^6}{82944 \pi^6}
+ \frac{5\lambda H^6}{512 \pi^6} \Big[ -\frac{2}{3 \epsilon}  +  \ln\big( \frac{H^2}{4 \pi\mu^2}  \big) + \frac{127}{54} - \gamma \Big]  \big( \xi -\frac{1}{6}  \big)
+{\cal O} \big( \big( \xi -\frac{1}{6}  \big)^2 \big) \,   \quad (n=6) .
\ea
From Eq.  (\ref{w-eq}) for $w$ in conformal point with
$n=6$ we obtain $w=-\frac{1}{3}$, so we expect that the singular terms in $\Delta \langle \rho \rangle $ and  $\Delta \langle P \rangle $ to be related with this equation of state. Indeed, we have
\ba
\Delta \langle \rho \rangle + 3 \Delta \langle P \rangle \simeq  \frac{35 \lambda H^6}{1152 \pi^6}  \big( \xi -\frac{1}{6}  \big) \quad  \quad \quad (n=6) \, ,
\ea
so the singular terms $\epsilon^{-1}$ are cancelled in the above expression.
We have checked that the equation of state $w=-\frac{1}{3}$ also holds for the most singular terms $\epsilon^{-2}$  which appears at second order  $\big( \xi -\frac{1}{6}  \big)^2 $.

A conclusion is that the equation of state  Eq. (\ref{w-eq}), which is obtained for the conformal point,  is the  equation of state for the singular terms in inverse powers of $\epsilon $ in $\langle \rho \rangle$ and $\langle P \rangle$ at each order of $ \big( \xi -\frac{1}{6}  \big)$ as well.

\subsection{Numerical plots for general value of $\xi$ }

As the analytical expressions for $\Delta \langle \rho \rangle $ and  $\Delta \langle T \rangle $ for the general values of $\xi$ are very complicated, here we present their  numerical plots for the special case of $n=4$.

\begin{figure}[h]
	\centering
\includegraphics[  width=0.45\linewidth]{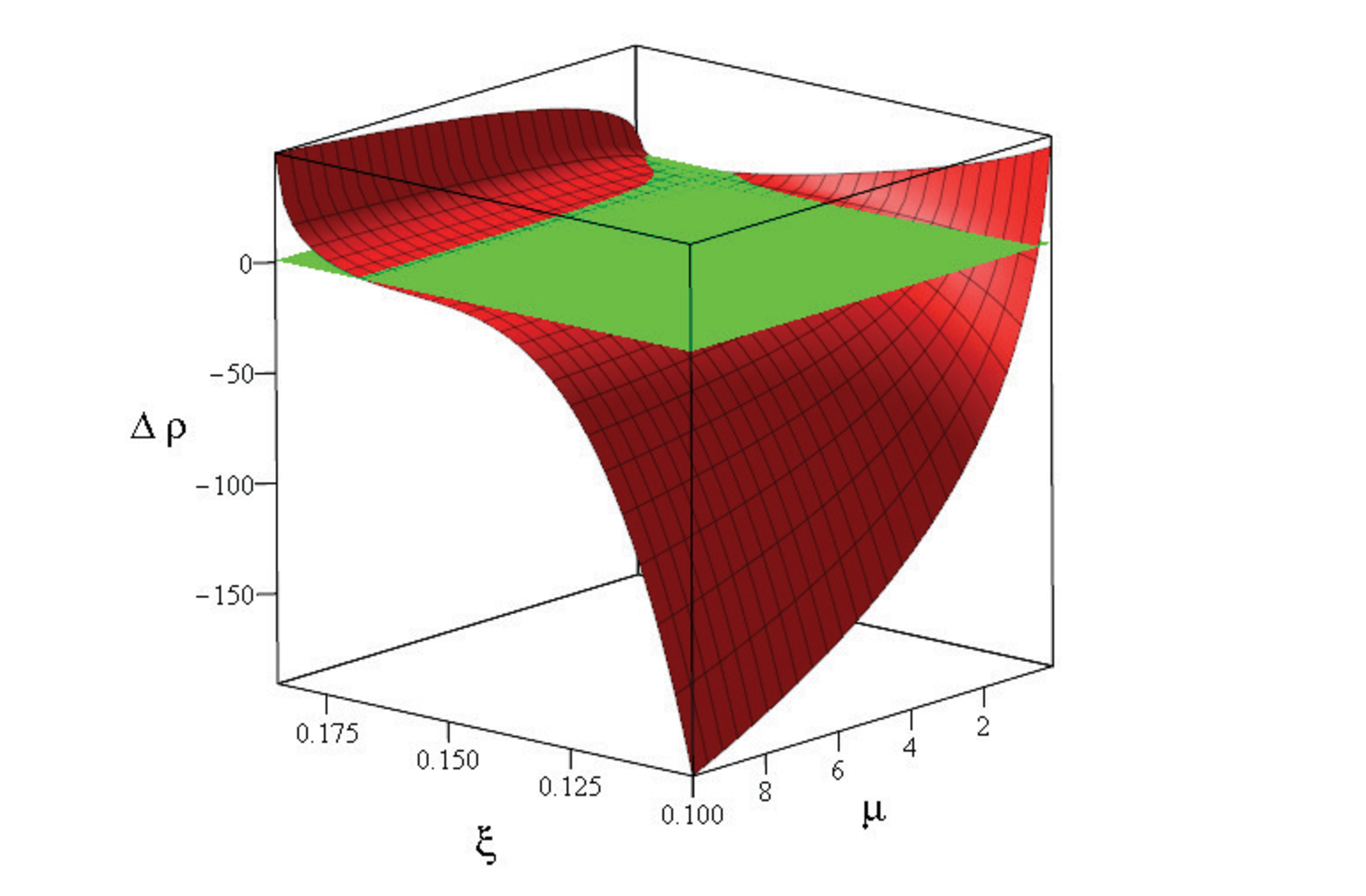}
\includegraphics[ width=0.54\linewidth]{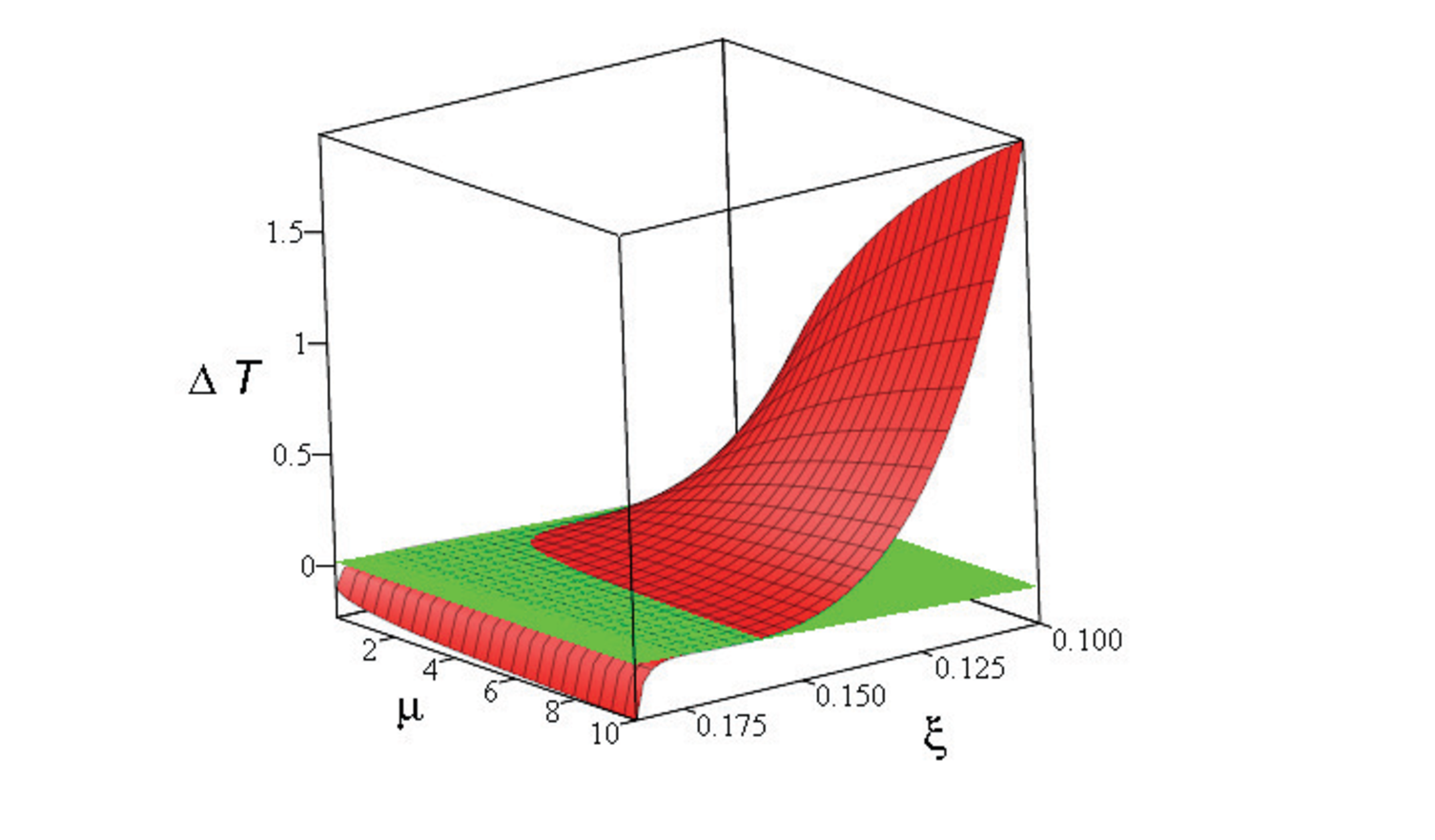}
	\vspace{-0.5cm}
	\caption{The 3D diagrams  of $\Delta \langle \rho \rangle$ and $\Delta \langle T \rangle$ as functions of variables $(\xi, \mu)$ for n=4. We have varied $\mu$ in units of $H$ while  $\xi$ is varied in the interval $0< \xi< \frac{3}{16}$. In the left panel $\Delta \langle \rho \rangle$ is measured in the scale of $\frac{-\lambda H^4}{2304 \pi^4}$ while in the right panel $\Delta \langle T \rangle$ is measured in the scale of
$ -\frac{3\lambda H^4}{16 \pi^4}$. The green horizontal surface in the left panel represents the surface $z=1$ while in the right panel it represents
the surface $z=\xi-\frac{1}{6}$. In the left panel near the conformal point $\xi=\frac{1}{6}$,  $\Delta \langle \rho \rangle$ approaches the constant value given in Eq. (\ref{Deltarho-n4}) while in the right panel
$\Delta \langle T\rangle$ approaches the formula given in Eq. (\ref{DeltaTrace-n4})  the near the conformal point.
}
\vspace{1cm}
\label{fig4}
\end{figure}

In Fig. \ref{fig4}, we have presented the three-dimensional behaviour of $\Delta \langle \rho \rangle $ and  $\Delta \langle T \rangle $ as functions of two parameters $(\xi, \mu)$. We have  varied $\mu$ in units of $H$ while
$\xi$ is varied in the interval $0< \xi< \frac{3}{16}$ in which the index $\nu$ is real. In the left panel of this figure, we have presented
$\Delta \langle \rho \rangle /(\frac{-\lambda H^4}{2304 \pi^4})$ while in the right panel we have presented $\Delta \langle T\rangle/( -\frac{3\lambda H^4}{16 \pi^4})$. In the left panel, the horizontal surface represents the surface with the value equal to unity while in the right panel the horizontal surface
represents the surface   $z=\xi-\frac{1}{6}$, independent of $\mu$.  In the left panel, we see that near the conformal point $\xi=\frac{1}{6}$,  $\Delta \langle \rho \rangle$ approaches the constant value given in Eq. (\ref{Deltarho-n4}). On the other hand, in the right panel we see that near the conformal point
$\Delta \langle T\rangle$ approaches the formula given in Eq. (\ref{DeltaTrace-n4}) in which the quantum corrections in trace vanish
at the conformal point.

\begin{figure}[t]
	\centering
\includegraphics[  width=0.5\linewidth]{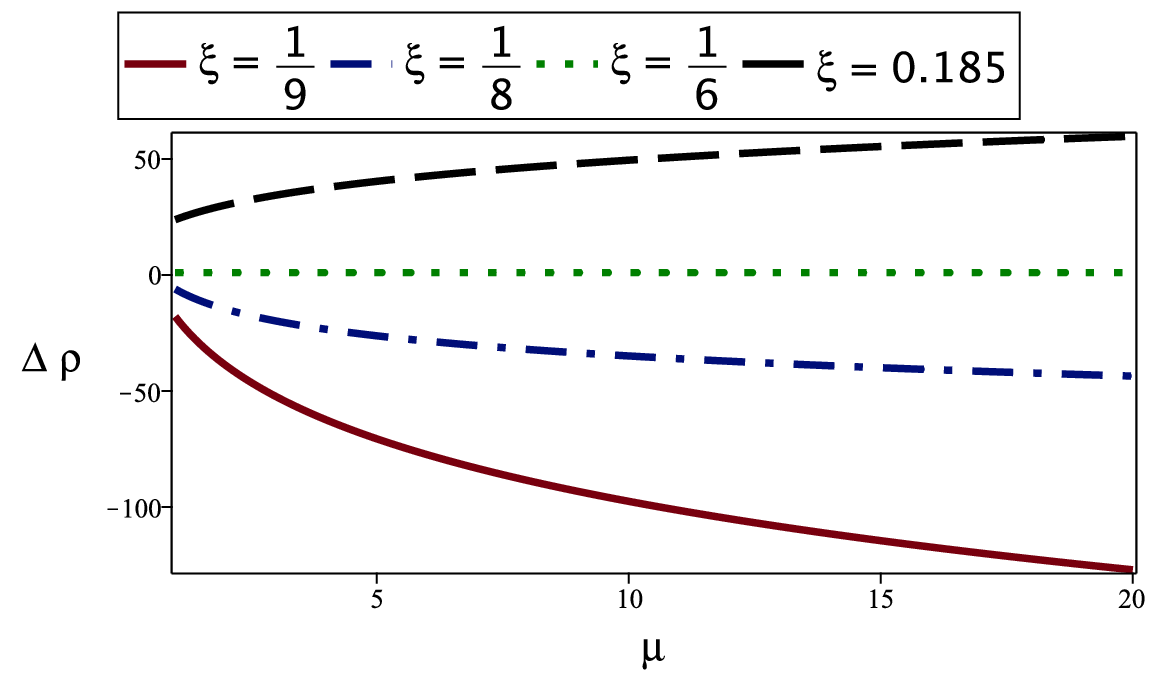}
\hspace{0.5cm}
\includegraphics[ width=0.43\linewidth]{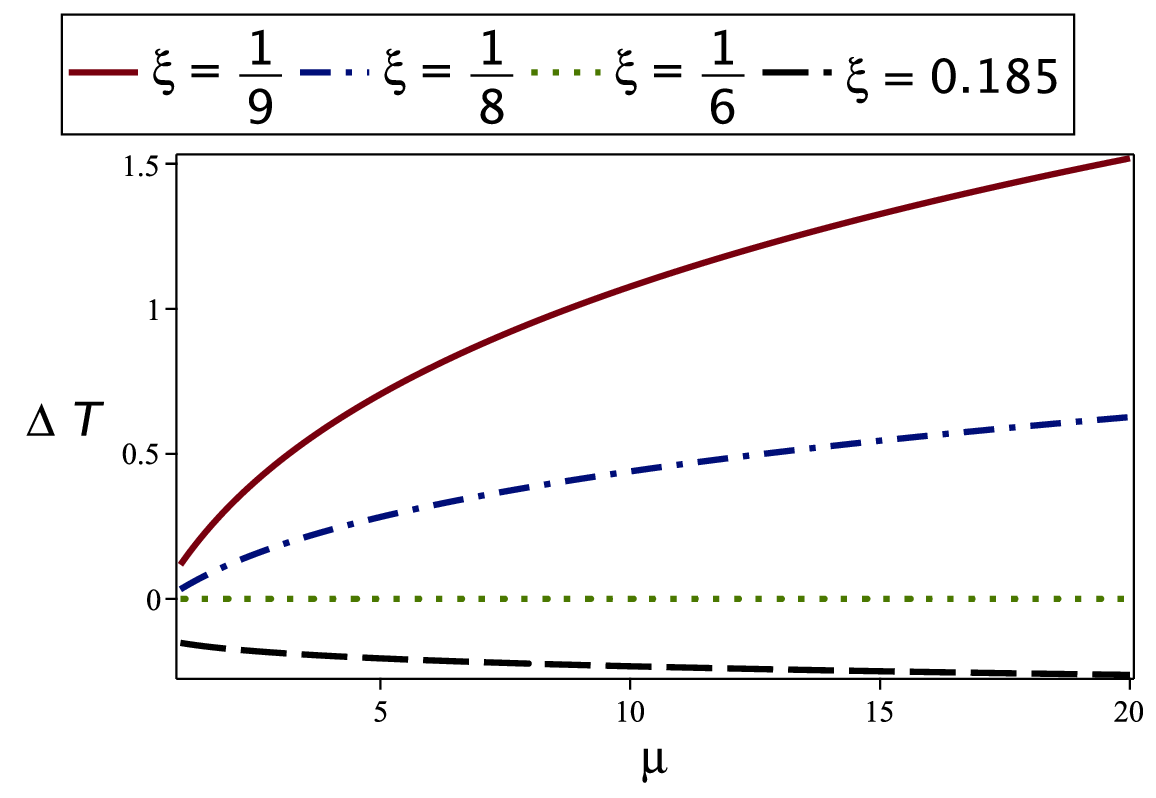}
	\caption{The diagram shows the behaviour of $\Delta \langle \rho \rangle$  and $\Delta \langle T \rangle$ as functions of $\mu$ for various fixed values of $\xi$. As in Fig. \ref{fig4}, $\Delta \langle \rho \rangle$ is measured in the scale of $\frac{-\lambda H^4}{2304 \pi^4}$ while  $\Delta \langle T \rangle$ is measured in the scale of $ \frac{-3\lambda H^4}{16 \pi^4}$. At the conformal point  $\xi= \frac{1}{6}$, $\Delta \langle \rho \rangle$  and $\Delta \langle T \rangle$ are constant while the behaviours of the curves change
for values of $\xi$ below and above this value.
}
\vspace{0.5cm}
\label{fig2}
\end{figure}

\begin{figure}[t]
	\centering
\includegraphics[  width=0.47\linewidth]{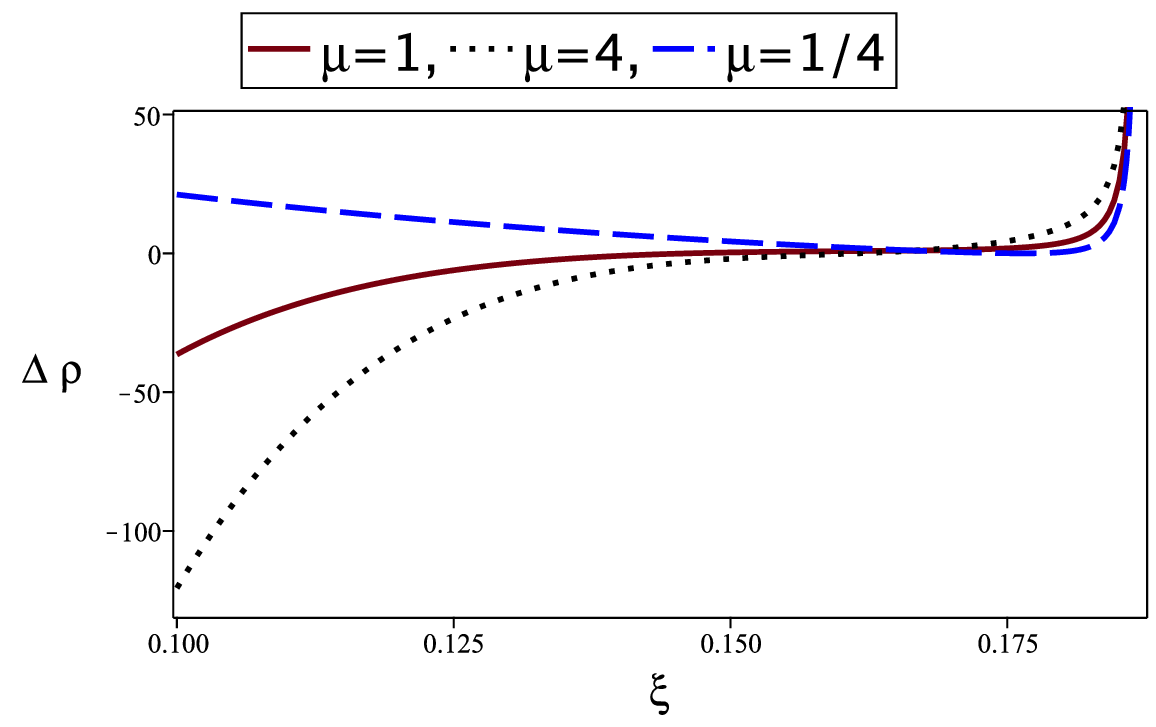}\hspace{0.3cm}
\includegraphics[ width=0.47\linewidth]{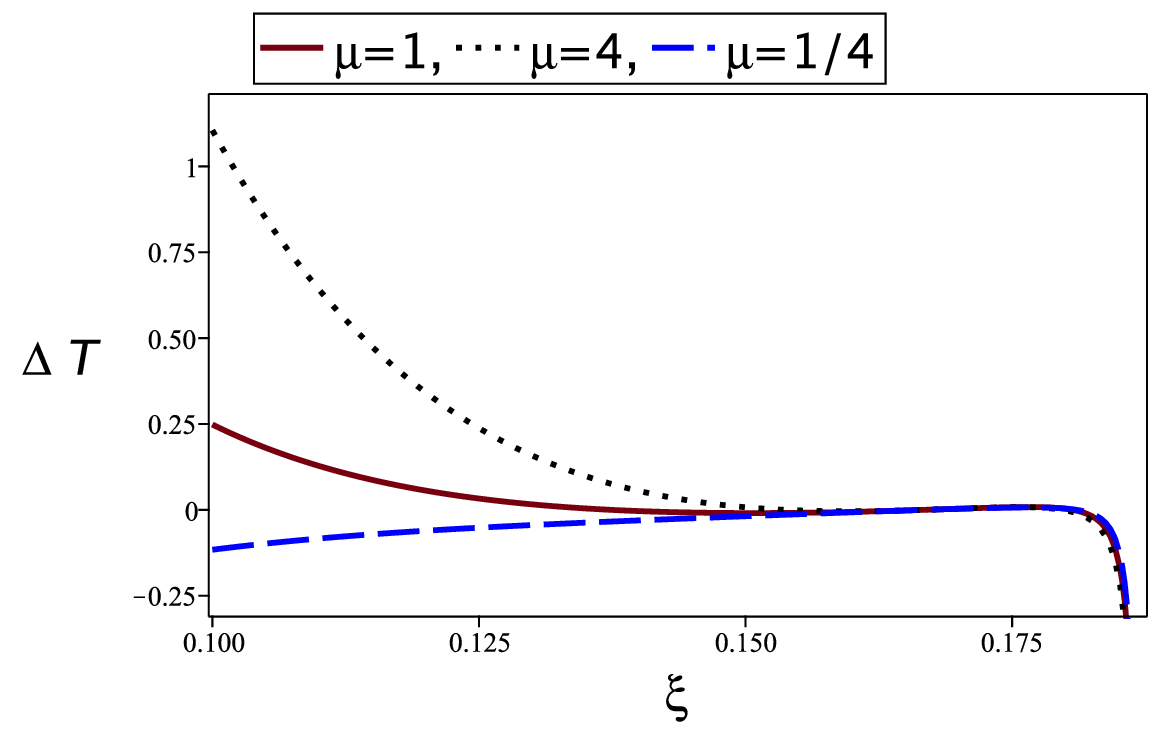}
	\caption{The diagram shows the behaviour of $\Delta \langle \rho \rangle$ and $\Delta \langle T \rangle$ as functions of $\mu$ for various fixed values of $\xi$. As in previous plots  $\Delta \langle \rho \rangle$ is measured in the scale of $\frac{-\lambda H^4}{2304 \pi^4}$ while  $\Delta \langle T \rangle$ is measured in the scale of $ \frac{-3\lambda H^4}{16 \pi^4}$. At the conformal point $\xi=\frac{1}{6}$, all curves merge to a fixed value in each panel.
}
\vspace{0.5cm}
\label{fig3}
\end{figure}

To have a better visualization, in Figs. \ref{fig2} and \ref{fig3} we have presented the two-dimensional sections of the above plot  as functions of $\mu$ ($\xi$) for fixed values of $\xi$ ($\mu$).   In both panels of Fig. \ref{fig2} we see that $\xi=\frac{1}{6}$ plays special role in which at this point
 $\Delta \langle \rho \rangle= \frac{-\lambda H^4}{2304 \pi^4} $ and  $\Delta \langle T \rangle =0$ independent of the value of $\mu$.   This property is reinforced in Fig. \ref{fig3} in which all curves merge to each other at the conformal point $\xi=\frac{1}{6}$.

\section{Summary and Discussions}

In this work we have studied the quantum fields with self-interaction potential $V \propto \lambda \Phi^n$ in a dS background and calculated the vacuum expectation values of the energy density, pressure and the trace of the energy momentum tensor. We have employed the perturbative in-in formalism to calculate the corrections in $\langle \rho \rangle_\Omega$ etc to first order in $\lambda$. To simplify the analysis we have considered even values of $n$. This is because for even values of $n$ the non-zero corrections appear at the first order of $\lambda$ while for odd values of
$n$ the  non-zero corrections appear at the  order $\lambda^2$. Technically speaking, this corresponds to having a single time in-in integral for even values of
$n$ while for odd values of $n$, we need to  consider nested (double time) in-in integrals.

Our analysis were performed  for  general values of $\xi$ but to report the analytical results we have considered two special limits, the conformal point $\xi =\frac{1}{6}$ and the case with a small deviation from the conformal point $\delta \xi \ll 1$. At the conformal point we obtain the equation of state  Eq.  (\ref{w-eq}) between
$\langle \rho \rangle_\Omega$ and $\langle P \rangle_\Omega$. In particular, for
$n=4$ we obtain the intriguing result that $w=\frac{1}{3}$ with $\langle T \rangle_\Omega =0$. This suggests that the classical conformal invariance associated to the case $n=4$ is respected under two-loop
(i.e. order $\lambda$) quantum corrections. It is an open question if this symmetry does hold to higher orders loops corrections. On the other hand, for other values of $n$, the coupling  $\lambda$ is dimensionful so the theory is not conformally invariant even at the classical level. Therefore, the equation of state is different that that of radiation. It approaches $w \rightarrow -\frac{2}{3}$ for large values of $n$.
In the case of $\delta \xi \neq 0$ but being small, we have calculated  $\langle \rho \rangle_\Omega$ and $\langle P \rangle_\Omega$ to first order in $\delta \xi$ for cases $n=4$ and $n=6$. Unlike the conformal point, we have the divergent terms
$\epsilon^{-1}$ at the order $\delta \xi$ and $\epsilon^{-2}$ for $n=6$ at the order $\delta \xi^2$. We have found that  these singular terms between $\langle \rho \rangle_\Omega$ and $\langle P \rangle_\Omega$ are related by the equation of state Eq.  (\ref{w-eq}) at each order of $\delta \xi$.

While we presented the analytical results only for the above two special cases, but  we have presented the numerical plots of $\Delta \langle \rho \rangle$ and $\Delta \langle T \rangle$ for general values of $\xi$ for $n=4$ in Figs. \ref{fig4}, \ref{fig2} and \ref{fig3}. All these figures highlight the special roles played by the conformal
limit $\xi= \frac{1}{6}$.

There are a number of directions in which the current study can be extended. The first which comes to mind is repeating this investigation for odd values of $n$. In this case, the in-in analysis become more complicated as one has to go to second order in in-in integrals, with corrections appearing at order $\lambda^2$. The second question is to extend the current analysis for $n=4$ to second order in $\lambda$ and see if the classical conformal invariance with $w=\frac{1}{3}$ and $\langle T \rangle_\Omega =0$ hold at order $\lambda^2$ or not. Since we were mostly interested in conformal limits, we have set the mass of the field to be zero. However, one can easily incorporate the effects of mass in the current analysis as well. In particular, as a physical example, one may consider a symmetry breaking potential like $ V = \lambda ( \Phi^2 - v^2)^2$ in which the effective mass, the cubic and the quartic couplings are all related via the coupling constant $\lambda$.  Employing in-in formalism to second order in $\lambda$, one can calculate the quantum corrections in vacuum zero point energy from both the cubic and quartic self-interactions. As the theory is not scale invariant even at the classical level, then one may not expect that the conclusions  $w=\frac{1}{3}$ and $\langle T \rangle_\Omega =0$ to hold under loop corrections. We would like to come back to the interesting example of symmetry breaking potential  in future.

\vspace{1cm}

 {\bf Acknowledgments:}  We are grateful to   Richard Woodard and Ali Akbar Abolhasani  for useful  discussions and correspondences. The work of H. F. is partially supported by INSF  of Iran under the grant  number 4022911. H.S. expresses gratitude to E. T. Akhmedov, D. V. Bykov, A. Bazarov, and I. Aref'eva for the constructive discussions they had during conference fields\&Strings 2024, Moscow, and appreciates the warm hospitality of the Steklov Institute.

\vspace{1cm}
\appendix

\section{Wick Contractions}
\label{contractions}

In this Appendix we briefly review some steps concerning
Wick contractions.

Let us consider  Eq. \eqref{rho3-int1} in configuration space,
\begin{equation}\label{Deltarho3-app}
\Delta \langle {\rho _3}(\tau ,{ {\mathbf{x}}_{\mathbf{0}} })\rangle  = \frac{{\lambda {H^2}}}{n}\operatorname{Im} \Big[\int_{ - \infty }^\tau  d \tau 'a{(\tau ')^D}\int {{d^{D - 1}}} x{\langle 0|\Phi {(\tau ,{{\mathbf{x}}_{\mathbf{0}}})^2}\Phi {(\tau ',{\mathbf{x}})^n}|0\rangle _0}\Big]\,.
\end{equation}
Going to Fourier space, and setting ${\mathbf{x}}_{\mathbf{0}}=0$ because of the translation invariance,  this yields
\begin{eqnarray}
\label{expanded-forms}\nonumber
  \Delta \langle {\rho _3}(\tau ,{{\mathbf{x}}_{\mathbf{0}}})\rangle  &=& \langle 0|\frac{{{\mu ^{(n/2)(4 - D)}}}}{n}\lambda {H^2}{\mathrm{Im}} \Big[\int_{ - \infty }^\tau  d \tau 'a{(\tau ')^D}\int {\frac{{{d^{D - 1}}{\mathbf{p}_1}}}{{{{(2\pi )}^{D - 1}}}}} \frac{{{d^{D - 1}}{\mathbf{p}_2}}}{{{{(2\pi )}^{D - 1}}}} \frac{{{d^{D - 1}}{\mathbf{q}_1}}}{{{{(2\pi )}^{D - 1}}}}\cdots \frac{{{d^{D - 1}}{\mathbf{q}_n}}}{{{{(2\pi )}^{D - 1}}}}\\\nonumber
&\times& \int {{d^{D - 1}}} x \left[ {{a_{{{\mathbf{p}}_1}}}\phi _{{\mathbf{p}_1}}^{}(\tau ){} + a_{{{\mathbf{p}}_1}}^\dag \phi _{{\mathbf{p}_1}}^*(\tau )} \right]  \left[ {{a_{{{\mathbf{p}}_2}}}{\phi _{{\mathbf{p}_2}}}(\tau ) + a_{ {{\mathbf{p}}_2}}^\dag \phi _{{\mathbf{p}_2}}^*(\tau ) } \right]
\\\nonumber
& \times&  \left[ {{a_{{{\mathbf{q}}_1}}}\phi _{{\mathbf{q}_1}}^{}(\tau '){e^{i{{\mathbf{q}}_1} \cdot {\mathbf{x}}}} + a_{{{\mathbf{q}}_1}}^\dag \phi _{{\mathbf{q}_1}}^*(\tau '){e^{ - i{{\mathbf{q}}_1} \cdot {\mathbf{x}}}}} \right]
 \left[ {{a_{{{\mathbf{q}}_2}}}\phi _{{\mathbf{q}_2}}^{}(\tau '){e^{i{{\mathbf{q}}_2} \cdot {\mathbf{x}}}} + a_{{{\mathbf{q}}_2}}^\dag \phi _{{\mathbf{q}_2}}^*(\tau '){e^{ - i{{\mathbf{q}}_2} \cdot {\mathbf{x}}}}} \right] \\\nonumber
 &\times&  \left[ {{a_{{{\mathbf{q}}_3}}}\phi _{{\mathbf{q}_3}}^{}(\tau '){e^{i{{\mathbf{q}}_3} \cdot {\mathbf{x}}}} + a_{{{\mathbf{q}}_3}}^\dag \phi _{{\mathbf{q}_3}}^*(\tau '){e^{ - i{{\mathbf{q}}_3} \cdot {\mathbf{x}}}}} \right]
  \left[ {{a_{{{\mathbf{q}}_4}}}\phi _{{\mathbf{q}_4}}^{}(\tau '){e^{i{{\mathbf{q}}_4} \cdot {\mathbf{x}}}} + a_{{{\mathbf{q}}_4}}^\dag \phi _{{\mathbf{q}_4}}^*(\tau '){e^{ - i{{\mathbf{q}}_4} \cdot {\mathbf{x}}}}} \right] \\\nonumber
 &\times& \cdots \left[ {{a_{{{\mathbf{q}}_{n}}}}\phi _{{\mathbf{q}_{n}}}^{}(\tau '){e^{i{{\mathbf{q}}_{n}} \cdot {\mathbf{x}}}} + a_{{{\mathbf{q}}_{n}}}^\dag \phi _{{\mathbf{q}_{n}}}^*(\tau '){e^{ - i{{\mathbf{q}}_{n}} \cdot {\mathbf{x}}}}} \right]|0\rangle\Big]\,.\\
\end{eqnarray}

As a specific example consider $n=6$ where we present the Wick contractions for one case:
\begin{equation}
\begin{gathered}\label{laddering-expression}
  \Delta \left\langle {{\rho _3}\left( {\tau ,{{\mathbf{x}}_{\mathbf{0}}}} \right)}
  \right \rangle  \supset  \langle  0 \Big |\frac{{{\mu ^{(3)(4 - D)}}}}{6}\lambda {H^2}\operatorname{Im} \left[ {\int_{ - \infty }^\tau  d {\tau ^\prime }a{{\left( {{\tau ^\prime }} \right)}^D}\int {\frac{{{d^{D - 1}}{{\mathbf{p}}_1}}}{{{{(2\pi )}^{D - 1}}}}} \frac{{{d^{D - 1}}{{\mathbf{p}}_2}}}{{{{(2\pi )}^{D - 1}}}}\frac{{{d^{D - 1}}{{\mathbf{q}}_1}}}{{{{(2\pi )}^{D - 1}}}} \cdots \frac{{{d^{D - 1}}{{\mathbf{q}}_6}}}{{{{(2\pi )}^{D - 1}}}}} \right. \hfill \\
  \,\,\,\,\,\,\,\,\,\,\,\,\,\,\,\,\,\,\,\,\,\,\,\,\,\times \int {{d^{D - 1}}} x \left[ {{a_{{{\mathbf{p}}_1}}}{a_{{{\mathbf{p}}_2}}}{a_{{{\mathbf{q}}_1}}}{a_{{{\mathbf{q}}_2}}}a_{{{\mathbf{q}}_3}}^\dag a_{{{\mathbf{q}}_4}}^\dag a_{{{\mathbf{q}}_5}}^\dag a_{{{\mathbf{q}}_6}}^\dag } \right]\, \times \left[ { {e^{i({{\mathbf{q}}_1} + {{\mathbf{q}}_2} - {{\mathbf{q}}_3} - {{\mathbf{q}}_4} - {{\mathbf{q}}_5} - {{\mathbf{q}}_6}) \cdot {\mathbf{x}}}}} \right] \hfill \\
  \,\,\,\,\,\,\,\,\,\,\,\,\,\,\,\,\,\,\,\,\,\,\,\,\,\,\,\,\,\left[ {{\phi _{{{\mathbf{p}}_1}}}(\tau ){\phi _{{{\mathbf{p}}_2}}}(\tau ){\phi _{{{\mathbf{q}}_1}}}\left( {{\tau ^\prime }} \right){\phi _{{{\mathbf{q}}_2}}}\left( {{\tau ^\prime }} \right)\phi _{{{\mathbf{q}}_3}}^*\left( {{\tau ^\prime }} \right)\phi _{{{\mathbf{q}}_4}}^*\left( {{\tau ^\prime }} \right)\phi _{{{\mathbf{q}}_5}}^*\left( {{\tau ^\prime }} \right)\phi _{{{\mathbf{q}}_6}}^*\left( {{\tau ^\prime }} \right)} \right]\Bigg] \Big|0\rangle\,. \\
\end{gathered}
\end{equation}
Using  Eq. \eqref{Noncommutative} and noting that
\begin{equation}\label{delta-quantization}
\int_{} {{d^{D - 1}}} x\,{e^{i\left( {{{\mathbf{q}}_{\mathbf{i}}} \pm {{\mathbf{q}}_{\mathbf{j}}}} \right) \cdot {\mathbf{x}}}} = {(2\pi )^{D - 1}}{\delta ^{(D - 1)}}\left( {{{\mathbf{q}}_{\mathbf{i}}} \pm {{\mathbf{q}}_{\mathbf{j}}}} \right)\,,
\end{equation}
then Eq. \eqref{laddering-expression} is cast into the following form,
\begin{equation}
\label{laddering-expression-reduced}
\begin{gathered}
\Delta   \left\langle{{\rho _3}\left( {\tau ,{{\mathbf{x}}_{\mathbf{0}}}} \right)} \right\rangle  \supset  \langle 0 \Big|\frac{{{\mu ^{3(4 - D)}}}}{6}\lambda {H^2}\operatorname{Im} \left[ {\int_{ - \infty }^\tau  d {\tau ^\prime }a{{\left( {{\tau ^\prime }} \right)}^D}\int {\frac{{{d^{D - 1}}{{\mathbf{p}}_1}}}{{{{(2\pi )}^{D - 1}}}}} \frac{{{d^{D - 1}}{{\mathbf{p}}_2}}}{{{{(2\pi )}^{D - 1}}}}\frac{{{d^{D - 1}}{{\mathbf{q}}_1}}}{{{{(2\pi )}^{D - 1}}}} \cdots \frac{{{d^{D - 1}}{{\mathbf{q}}_6}}}{{{{(2\pi )}^{D - 1}}}}} \right. \hfill \\
  \,\,\,\,\,\,\,\,\,\,\,\,\,\,\,\,\,\,\,\,\,\,\ \delta ({{\mathbf{q}}_2} - {{\mathbf{q}}_3})\delta ({{\mathbf{q}}_1} - {{\mathbf{q}}_4})\delta ({{\mathbf{p}}_1} - {{\mathbf{q}}_5})\delta ({{\mathbf{p}}_2} - {{\mathbf{q}}_6})\delta ({{\mathbf{q}}_1} + {{\mathbf{q}}_2} - {{\mathbf{q}}_3} - {{\mathbf{q}}_4} - {{\mathbf{q}}_5} - {{\mathbf{q}}_6}) \hfill \\
 \,\,\,\,\,\,\,\,\,\,\,\,\,\,\,\,\,\,\,\,\,\,\  \left[ {{\phi _{{{\mathbf{p}}_1}}}(\tau ){\phi _{{{\mathbf{p}}_2}}}(\tau ){\phi _{{{\mathbf{q}}_1}}}\left( {{\tau ^\prime }} \right){\phi _{{{\mathbf{q}}_2}}}\left( {{\tau ^\prime }} \right)\phi _{{{\mathbf{q}}_3}}^*\left( {{\tau ^\prime }} \right)\phi _{{{\mathbf{q}}_4}}^*\left( {{\tau ^\prime }} \right)\phi _{{{\mathbf{q}}_5}}^*\left( {{\tau ^\prime }} \right)\phi _{{{\mathbf{q}}_6}}^*\left( {{\tau ^\prime }} \right)} \right]\Bigg] \Big|0\rangle\,.
\end{gathered}
\end{equation}

After some manipulations, the above equation  takes the following form
\begin{eqnarray}\label{delta-contractions}\nonumber
\begin{gathered}
  \left\langle {{\rho _3}\left( {\tau ,{{\mathbf{x}}_{\mathbf{0}}}} \right)} \right\rangle  \supset \frac{{{\mu ^{3(4 - D)}}}}{6}\lambda {H^2}\operatorname{Im} \left[ {\int_{ - \infty }^\tau  d {\tau ^\prime }a{{\left( {{\tau ^\prime }} \right)}^D}\int {\frac{{{d^{D - 1}}{{\mathbf{q}}_1}}}{{{{(2\pi )}^{D - 1}}}}} \frac{{{d^{D - 1}}{{\mathbf{q}}_2}}}{{{{(2\pi )}^{D - 1}}}}\frac{{{d^{D - 1}}{{\mathbf{q}}_5}}}{{{{(2\pi )}^{D - 1}}}}} \right. \\
  \left[ {{{\left| {{\phi _{{{\mathbf{q}}_1}}}\left( {{\tau ^\prime }} \right)} \right|}^2}{{\left| {{\phi _{{{\mathbf{q}}_2}}}\left( {{\tau ^\prime }} \right)} \right|}^2}\phi _{{{\mathbf{q}}_5}}^2(\tau )\phi _{{{\mathbf{q}}_5}}^{*2}\left( {{\tau ^\prime }} \right)} \right]\Bigg]\,.
\end{gathered}
\end{eqnarray}
It is easy to show that  the desired equation for  $\Delta \langle {\rho _3}(\tau ,{{\mathbf{x}}_{\mathbf{0}}})\rangle$ i.e. Eq. \eqref{rho3-int2},  for $n=4$ is obtained accordingly.

{}

\end{document}